\documentclass[%
 aip,
 amsmath,amssymb,
 reprint,%
]{revtex4-1}

\usepackage[utf8]{inputenc}
\usepackage[OT4]{fontenc}
\usepackage{amsmath}
\usepackage{amssymb}
\usepackage{graphicx}
\usepackage{subcaption}
\usepackage{enumitem}
\usepackage{mathptmx}
\usepackage{etoolbox}

\DeclareMathOperator{\sign}{\mathrm{sgn}}
\DeclareMathOperator{\Tr}{\mathrm{Tr}}

\usepackage[colorlinks=true,hyperfootnotes=true,breaklinks=true]{hyperref}
\usepackage{cleveref}

\makeatletter
\def\@email#1#2{%
 \endgroup
 \patchcmd{\titleblock@produce}
  {\frontmatter@RRAPformat}
  {\frontmatter@RRAPformat{\produce@RRAP{*#1\href{mailto:#2}{#2}}}\frontmatter@RRAPformat}
  {}{}
}%
\makeatother

\begin{document}
\title{Thermal properties of structurally balanced systems on classical random graphs}

\affiliation{AGH University,
Faculty of Physics and Applied Computer Science,
al. Mickiewicza 30, 30-059 Krak\'ow, Poland}

\author{Krzysztof Malarz}
  \email{malarz@agh.edu.pl}
  \thanks{ORCID~\href{https://orcid.org/0000-0001-9980-0363}{0000-0001-9980-0363}}

\author{Maciej Wo{\l}oszyn}
  \thanks{ORCID~\href{https://orcid.org/0000-0001-9896-1018}{0000-0001-9896-1018}}

\date{\today}

\begin{abstract}
    The dynamics of social relations and the possibility of reaching the state of structural balance (Heider balance) under the influence of the temperature modeling the social noise level are discussed for interacting actors occupying nodes of classical random graphs.
    Depending on the graph density $D$, either a smooth cross-over or a first-order phase transition from a balanced to an imbalanced state of the system is observed with an increase in the thermal noise level.
    The minimal graph density $D_\text{min}$ for which the first-order phase transition can be observed decreases with the system size $N$ as $D_\text{min}\propto N^{-0.58(1)}$.
    For graph densities $D>D_\text{min}$ the reduced critical temperature $T_c^\star=T_c/T_c(D=1)$ increases with the graph density as $T_c^\star\propto D^{1.719(6)}$ independently of the system size $N$.  
\end{abstract}

\maketitle

\begin{quotation}
Modeling social processes gives us a unique opportunity to understand the dynamics of relations between individuals or larger communities.
Among the most studied phenomena are the processes governing changes in social opinion, while another example is the evolving structure of friendly and hostile relations.
Our work is devoted to the latter problem, which we analyze in terms of the so-called Heider balance (also known as structural balance), based on relations in triads of actors in which the cognitive dissonance may be easily associated with the imbalanced state.

Clearly, one of the most important factors is the structure of the network that connects these actors.
Here, we study the case of networks modeled with classical random graphs with varying graph density affecting the number of interacting triads of actors.
We discuss the influence of the noise introduced in terms of temperature variable, the possibility of reaching a balanced state, and phase transitions between the balanced and imbalanced states.

Our results show that above a certain minimal graph density, a phase transition of the first kind is observed, with the critical temperature dependent on the system size.
Above the minimal density required for the phase transition, a gradual and smooth crossover between balanced and imbalanced states occurs.
The presence of the phase transition is directly related to the number of clusters of triads that may interact (at least indirectly) during time evolution: we show that it is possible only when the system consists of a large enough number of triads organized into just one or a small number of such clusters.
\end{quotation}

\section{Introduction}

\begin{figure}[htbp!]
\begin{subfigure}{.240\columnwidth}
\centering
\caption{\label{fig:1a}}
\includegraphics[width=0.7\textwidth]{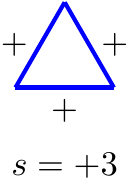}
\end{subfigure}
\begin{subfigure}{.240\columnwidth}
\centering
\caption{\label{fig:1b}}
\includegraphics[width=0.7\textwidth]{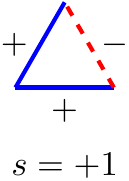}
\end{subfigure}
\begin{subfigure}{.240\columnwidth}
\centering
\caption{\label{fig:1c}}
\includegraphics[width=0.7\textwidth]{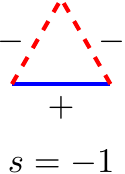}
\end{subfigure}
\begin{subfigure}{.240\columnwidth}
\centering
\caption{\label{fig:1d}}
\includegraphics[width=0.7\textwidth]{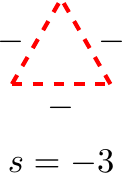}
\end{subfigure}
\caption{\label{fig:triads}(Color online). Heider's triads corresponding to balanced [(a) and (c)] and imbalanced [(b) and (d)] states.
Solid blue lines and dashed red lines represent friendly (positive, $x_{ij}=+1$) and hostile (negative, $x_{ij}=-1$) relations, respectively.
The values of $s=x_{ij}+x_{jk}+x_{ki}$ are presented in the bottom of subfigures and $i$, $j$, $k$ are the triangles vertices labels.}
\end{figure}

The signed network with positive and negative link values $x_{ij}=\pm 1$ representing friendly and hostile relations among actors---occupying network nodes $i$ and $j$---is termed structurally balanced \cite{Harary_1953,*Cartwright_1956,*Harary_1959,*Davis_1967,*Harary} when in every triangle these relations obey Heider's rules \cite{Heider}: 
\begin{itemize}[noitemsep]
\item a friend of my friend is my friend,
\item a friend of my enemy is my enemy,
\item an enemy of my friend is my enemy,
\item an enemy of my enemy is my friend.
\end{itemize}
The triangles where these rules are not violated are termed balanced (in the Heider sense).
The appearance of (imbalanced) triangles that do not obey these rules leads to the appearance of mental stress known as cognitive dissonance \cite{Festinger_1957,*Festinger_1962}.
Four possible signed triangles are shown in \Cref{fig:triads}. 
Among them, those with the sum $s=x_{ij}+x_{jk}+x_{ki}$ equal to $+3$ and $-1$ are balanced, while two others are not and $i$, $j$, $k$ are the labels of the triangle vertices. 
 
Several techniques are applied for modeling the resolving of the stress mentioned above, in which the imbalanced triangles are subsequently eliminated, leading the system towards the Heider balance (see Ref.~\onlinecite{Belaza_2017} for review).
For example, with the cellular automaton technique \cite{2005.11402} and the triangular lattice, the deterministic rule
\begin{equation}
x_{ij}(t+1)=\sign\big(x_{im}(t)x_{jm}(t) + x_{in}(t)x_{jn}(t)\big)
\label{rule:ca}
\end{equation}
is applied synchronously to all links $x_{ij}$ (where $i$, $j$, $m$, and $n$ are node labels in a pair of triangles $ijm$ and $ijn$ having a common edge $ij$). 
For a complete graph \cite{PhysRevE.104.024307}, the rule \eqref{rule:ca} may be generalized as
\begin{equation}
x_{ij}(t+1)=\sign\left(\sum_k x_{ik}(t)x_{jk}(t)\right),
\label{rule:cg}
\end{equation}
where $k\ne i,j$ is the node label of the third vertex of the triangle. Finally, \Cref{rule:cg} can be further generalized to any network \cite{2106.03054} defined by its adjacency matrix $\mathbf{A}=[a_{ij}]$ as
\begin{equation}
x_{ij}(t+1)=\sign\left(\sum_k a_{ik}x_{ik}(t)\cdot a_{jk}x_{jk}(t)\right),
\label{rule:net}
\end{equation}
for existing edges (where $a_{ij}=1$).
The (binary and symmetric) adjacency matrix $\mathbf{A}=[a_{ij}]$ elements
\begin{equation}
a_{ij}=\begin{cases}
1 & \iff i \text{ and } j \text{ are connected},\\
0 & \iff \text{otherwise},
\end{cases}
\end{equation}
define the network.
In all cases [\Cref{rule:ca,rule:cg,rule:net}] if the argument for the function $\sign(\cdots)$ is zero, $x_{ij}$ remains unchanged. 

The time evolution of the system may also be governed by a set of differential equations \cite{ISI:000362055900008,ISI:000357119200001,Gawronski,Gawronski_2005b,Marvel_2011,Kulakowski2005,Stojkow_2020}
\begin{equation}
\dot x_{ij}=\sum_k x_{ik}x_{jk}, 
\label{rule:sde}
\end{equation}
where $x_{ij}$ denotes the (real) value on the edge between the nodes $i$ and $j$ and the summation passes through all the nodes $k$ that make up the triangles $ijk$.
This equation is solvable analytically; however, $x_{ij}$ tends to infinity in a finite time \cite{Marvel_2011}.
To avoid this problem, the solution is either to impose numerically the condition $|x_{ij}|<1$, or to use the expression $(1-x_{ij}^2)$ as a prefactor \cite{Kulakowski2005}.
With this modification, \Cref{rule:sde} generically drives the system to one of $2^{N-1}$ balanced states. Other stationary solutions (the so-called jammed states) are also possible \cite{Stojkow_2020}, similarly to the discrete evolution \cite{Antal_2005}.

For modeling the level of social noise (equivalent to the non-zero temperature \cite{Microsociology,Macrosociology,social_temperature,1902.03454,*2002.05451,*2211.04183}),
the deterministic rules [\Cref{rule:ca,rule:cg,rule:net,rule:sde}] should be replaced by the probabilistic ones.
Examples of such rules (governed by the heat-bath algorithm \cite{PhysRevB.33.7861,Loison_2004}) were applied to the linear chain of actors \cite{2008.06362}, the triangular lattice \cite{2007.02128}, diluted and densified triangulations \cite{2106.03054} and complete graphs \cite{PhysRevE.99.062302,PhysRevE.100.022303,2009.10136,2011.07501}.
In the latter case, approximated mean-field solutions are also available \cite{PhysRevE.99.062302,2008.00537,1911.13048}.
In the presence of thermal noise, the system undergoes a phase transition from a balanced (low noise level) to an imbalanced (high noise level) state.

Recently, the absence of the above-mentioned phase transition was observed for the triangular lattice \cite{2007.02128}: independently of the noise level, the system remains in an imbalanced phase.
However, the signatures of various types of phase transitions were also described in earlier works on structurally balanced systems \cite{2008.06362,2010.10036}.

Intrigued by those observations, we checked the thermal evolution of the system for diluted and enriched triangular lattices \cite{2106.03054}.
We found that both the balanced (or partially balanced) and imbalanced states are possible if the average node degree is far enough from that of the regular triangular lattice.
The former state is observed at lower temperatures (noise levels) and the latter at higher temperatures.
Diluted triangular lattices and less enhanced networks show a smooth cross-over between those two states, and a phase transition of the first kind is observed for graphs of sufficiently high density.
In enhanced triangular lattices, the temperatures of the crossover as well as critical temperatures of the phase transition depend on the system size, whereas the crossover temperatures in diluted triangular lattices are size-independent.
If lattices are created by adding or removing only a small fraction of links to or from the regular triangular lattice, balanced states are not possible.

In this paper, we take a step further in investigating this issue and check the influence of thermal noise on structural balance in classic random graphs \cite{ER-1959,ER-1960,Random_Graphs_2001,Introduction_to_Random_Graphs_2015}.
We decided to use the heat-bath algorithm to be consistent with our earlier studies on triangular lattices \cite{2007.02128,2106.03054} and complete graph \cite{2009.10136,2011.07501}.
But in literature known to us, the alternatives---for instance Metropolis algorithm---are also applied (see for instance Refs.~\onlinecite{PhysRevE.99.062302,Pham_2020,Pham_2022}).
Finally, as we mentioned above, the heat-bath scheme of links update for complete graph is tractable also analytically in mean-field approximation and reaches perfect agreement with computer simulations realized with heat-bath algorithm \cite{1911.13048,2009.10136}.

\section{Model}

When the evolution of the system is governed by the heat-bath algorithm \cite{PhysRevB.33.7861,Loison_2004} (see also Ref.~\onlinecite[p.~154]{Guide_to_Monte_Carlo_Simulations_2009}) the deterministic rules [\Cref{rule:ca,rule:cg,rule:net,rule:sde}] should be replaced by the probabilistic ones,
\begin{subequations}
\label{eq:evol}
\begin{equation}
\label{eq:evol_x}
x_{ij}(t+1)=
        \begin{cases}
	+1 & \text{ with probability }p_{ij}(t),\\
	-1 & \text{ with probability }[1-p_{ij}(t)],
        \end{cases}
\end{equation}
where
\begin{equation}
\label{eq:evol_p}
    p_{ij}(t)=\frac{\exp[\xi_{ij}(t)/T]}{\exp[\xi_{ij}(t)/T]+\exp[-\xi_{ij}(t)/T]}.
\end{equation}
$T$ is the temperature (noise level) at which the evolution occurs, and
\begin{equation}
\label{eq:evol_xi}
	\xi_{ij}(t)=\sum_{k} a_{ik} x_{ik}(t) \cdot a_{kj} x_{kj}(t).
\end{equation}
\end{subequations}

Classical random graphs can be formed by a random combination of $N$ nodes with $L$ edges (the Erd{\H o}s--R\'enyi construction \cite{ER-1959,ER-1960}), or by implementing each of the possible $N(N-1)/2$ connections between nodes with a specified probability $D$ (the Gilbert construction \cite{Gilbert_1959}). In the thermodynamic limit ($N\to\infty$) both, the Gilbert and Erd{\H o}s--R\'enyi constructions lead to the same result ($D=2L/[N(N-1)]$) \cite{Kulakowski2005a}.
Here, we apply the Gilbert approach and term the probability $D$ as {\em a graph density}.
Then, every element of the adjacency matrix $a_{ij}=a_{ji}$ ($i\ne j$) is set to 1 with probability $D$ or 0 with probability $(1-D)$.

The Heider balance can be easily identified by checking the system work function \cite{Antal_2005,Krawczyk_2017}
\begin{equation}
\label{eq:Udef}
U\equiv-\frac{\Tr[(\mathbf{A}\circ\mathbf{X})^3]}{\Tr(\mathbf{A}^3)},
\end{equation}
where $\circ$ stands for the Hadamard product of the matrices and the matrix $\mathbf{X}=[x_{ij}]$.
The system work function $U$ is equal to $-1$ if and only if all triangles in the system are balanced.

Finally, to check the presence and kind of phase transition, we use the fourth-order Binder cumulant $K$ (see Ref.~\onlinecite[p.~78]{Guide_to_Monte_Carlo_Simulations_2009}) of the work function $U$ defined as
\begin{equation}
\label{eq:K}
K\equiv 1-\frac{\langle U^4 \rangle}{3 \langle U^2 \rangle^2},
\end{equation}
where $\langle \cdots \rangle$ denotes averaging over independent simulations.

The type of phase transition can be determined for a particular network based on the details of Binder cumulant dependence on temperature:
\begin{itemize}
    \item the first order (abrupt) phase transition is indicated by a deep minimum of $K(T)$, which is observed in the vicinity of the critical temperature $T_{\text{c}}$ \cite{PhysRevB.34.1841,PhysRevE.59.218} (see also Ref.~\onlinecite[p.~85]{Guide_to_Monte_Carlo_Simulations_2009}),
    \item the second order (continuous) phase transition corresponds to $K$ changing smoothly from $K=2/3$ at low temperatures to $K=0$ at high temperatures \cite{PhysRevE.59.218} (see also Ref.~\onlinecite[p.~508]{Binder_1997}), with non-trivial behavior of $K(T_{\text{c}})$ for various system sizes (see Ref.~\onlinecite[p.~80]{Guide_to_Monte_Carlo_Simulations_2009}).
\end{itemize}

\section{Results\label{sec:results}}

Our simulations performed on classical random graphs cover the whole range of graph densities $D$ for which the possibility of structural balance can be analyzed.
We start from values of density close to zero (at least one triad is needed) and proceed to $D=1$ (which is simply the complete graph), considering system sizes varying between $N=50$ and $N=400$.
In all cases, the initial state is imbalanced ($U=0$), with each $x_{ij}$ randomly set to $-1$ or $+1$ with the same probability.
All results presented below were obtained by completing the $t_{\text{max}}=10^4$ time steps, of which the last $\tau = 10^3$ were taken into account during the averaging procedure to find the outcome of a simulation.
For each set of parameters (temperature $T$, graph density $D$, system size $N$) $R=100$ independent simulations were performed and the mean values of the required quantities were calculated.

\begin{figure*}
\includegraphics[width=\textwidth]{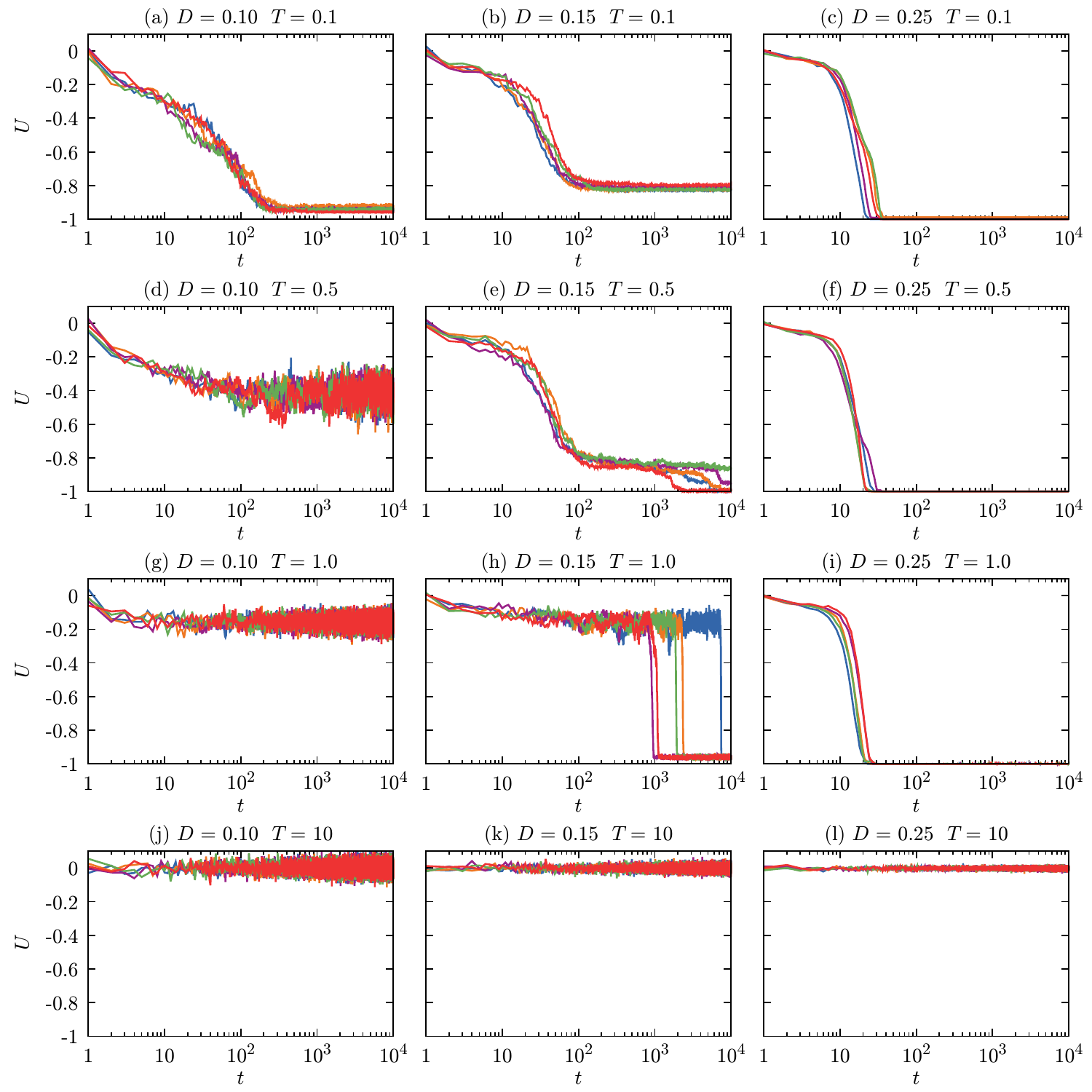}
\caption{\label{fig:crg-U_vs_time}(Color online). Time evolution of work function $U$ for systems described by classical random graphs. Results obtained in five independent simulations for $N=200$ and various 
densities, $D=0.10$, $0.15$, $0.25$, and temperatures $T=0.1$, $0.5$, $1$ and $10$.}
\end{figure*}

\begin{figure}
\includegraphics[width=\columnwidth]{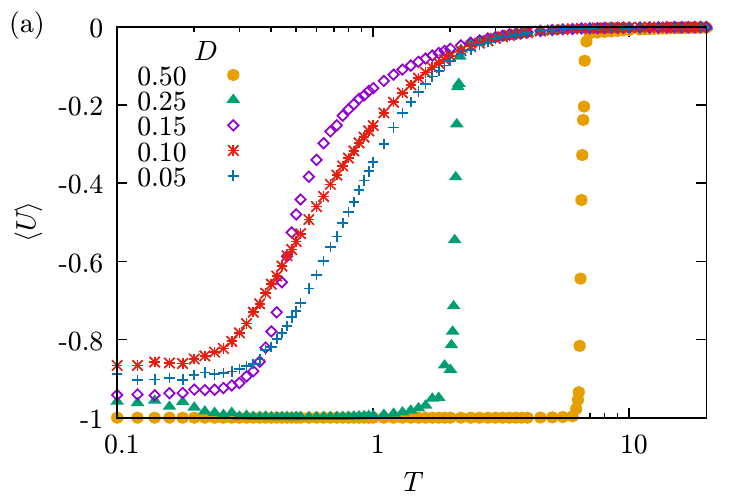}\\
\includegraphics[width=\columnwidth]{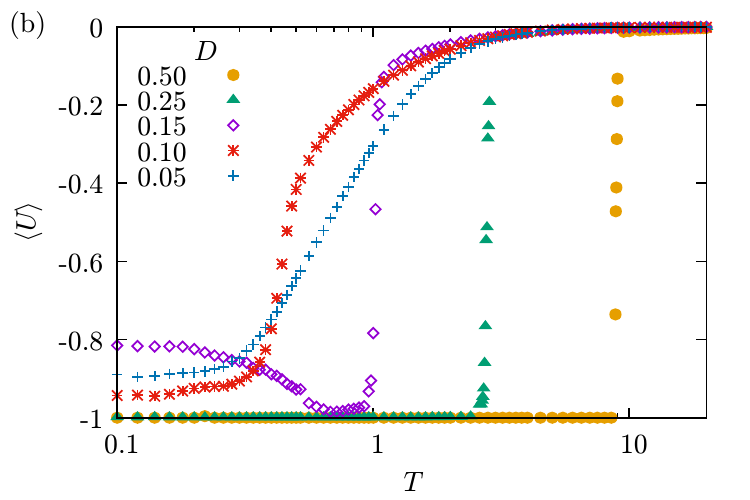}\\
\includegraphics[width=\columnwidth]{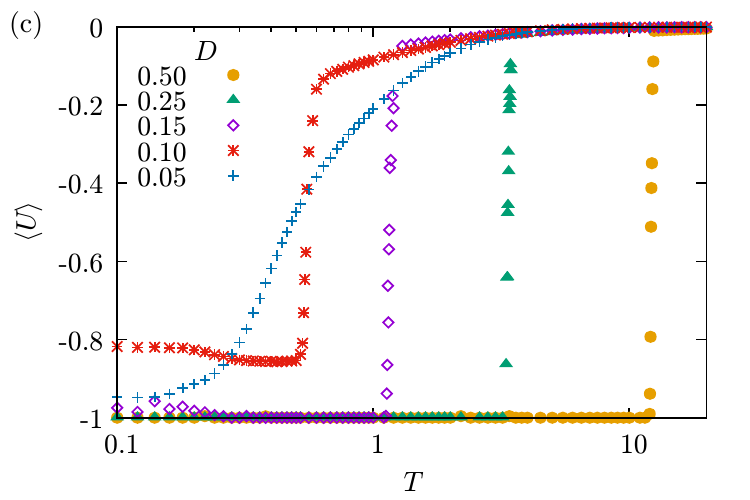}
\caption{\label{fig:crg-U_vs_T}(Color online). Thermal evolution of work function $\langle U\rangle$ for systems described by classical random graphs with
various density $D$ and sizes (a)~$N=100$, (b)~$N=200$, (c)~$N=400$.}
\end{figure}

The work function given by \Cref{eq:Udef}, which allows us to distinguish between balanced and imbalanced states, is calculated as a function of time for a range of temperatures.
Several examples of these time dependencies are presented in \Cref{fig:crg-U_vs_time}, with five simulations for each set of parameters.
It is clearly visible that the final state depends not only on the temperature $T$, but also on the density of the graph $D$.

\Cref{fig:crg-U_vs_T} shows the results of the calculations performed for several different densities, from relatively low $D=0.05$ to $D=0.5$, and as explained above averaged over $R$ simulations and $\tau$ time steps.
As expected, in the high-temperature limit the system is in an imbalanced ($\langle U \rangle = 0$) and disordered ($\{ x_{ij}\}=0$, where $\{ \cdots \}$ stands for averaging over all links) state.
On the other hand, in the opposite limit of $T \to 0$ the work function assumes values equal to or close to $-1$, which shows that at low temperatures the system reaches a balanced state or at least becomes partially balanced.

However, the change between the high- and low-temperature limits of $\langle U \rangle$ does not always occur in the same way.
At higher graph densities, the decrease in temperature causes the work function to abruptly change from $0$ to $-1$ at a certain temperature, whereas lower densities result in a smooth crossover from an imbalanced state toward the balanced state (not necessarily reaching the perfect balance).
Comparison of parts (a), (b), and (c) of \Cref{fig:crg-U_vs_T} also indicates that the size $N$ of the system plays a significant role, since for the same $D$ the characteristics of $\langle U(T) \rangle$ are different for various $N$.
In particular, the temperature at which the system becomes balanced is not the same.
For a complete graph ($D=1$) this dependence becomes $T_{\text{c}}\propto\sqrt{N}$ \cite{2009.10136}.

Furthermore, a more detailed study reveals that in some cases, a drop in $\langle U(T) \rangle$ resulting from the decrease in $T$ is followed by a slight increase observed at the lowest $T$. 
This effect is clearly visible for $N=200$ and $D=0.15$ shown in \Cref{fig:crg-U_vs_T}(b), where $\langle U \rangle \approx -1$ directly below $T_{\text{c}} \approx 1$, but when $T$ is even lower, the work function increases to $\langle U \rangle \approx -0.8$ when $T \to 0$.
The same applies, for example, to the case of $N=100$ and $D=0.25$ shown in \Cref{fig:crg-U_vs_T}(a), or $N=400$ and $D=0.1$ shown in \Cref{fig:crg-U_vs_T}(c), although in those two situations the increase of $\langle U \rangle$ when $T$ approaches zero is much smaller.
It is, however, most probably related to the time needed to reach the final state, which can become very long in some cases---as proved e.g. by the results shown in \Cref{fig:crg-U_vs_time}(e).
The possible reasons for this phenomenon, including the number of triads for which $\xi_{ij}=0$ when $D$ decreases,  will be discussed in \Cref{sec:discussion}.

\begin{figure}
    \includegraphics[width=0.99\columnwidth]{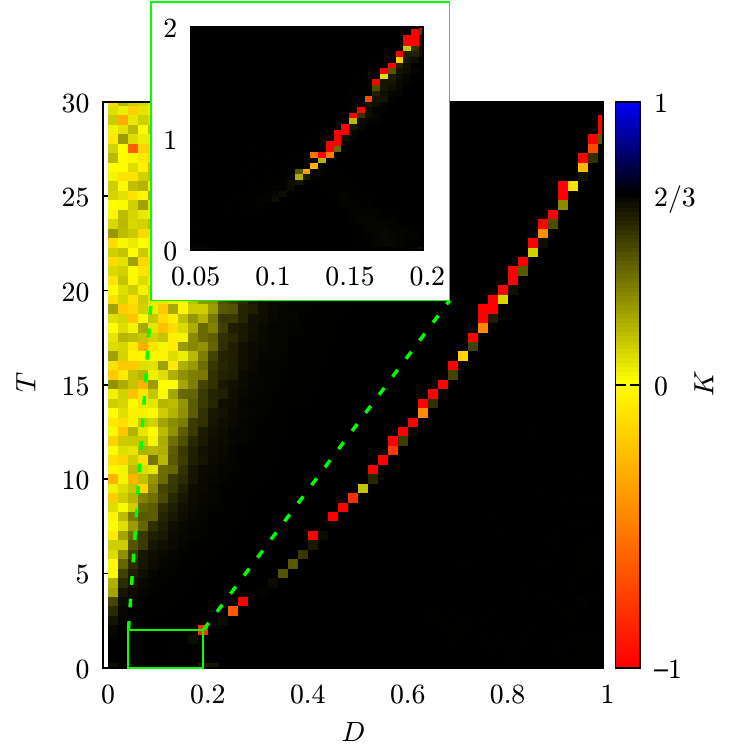}
    \caption{\label{fig:K-D-T}The fourth-order Binder cumulant $K$ as a function of graph density $D$ and temperature $T$ for $N=200$.}
\end{figure}

\begin{figure}
    \includegraphics[width=0.99\columnwidth]{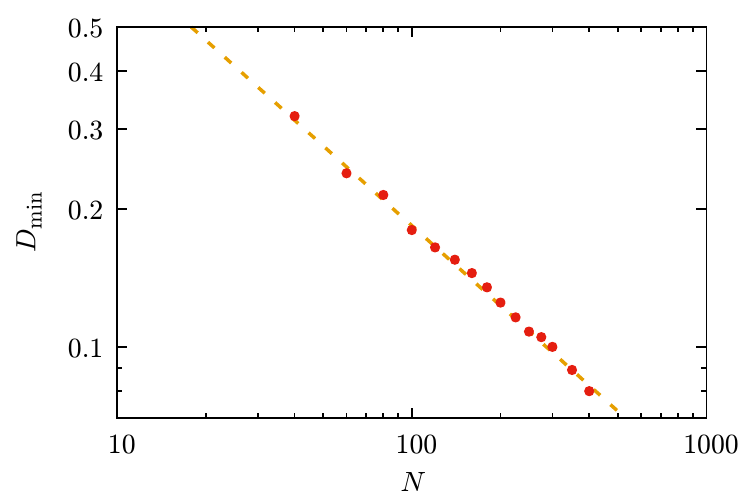}
    \caption{\label{fig:Dmin}The minimal density $D_{\text{min}}$ required for the phase transition between the balanced and imbalanced state. The dashed line is the result of least-squares fitting, $D_{\text{min}} = 2.7(1) \cdot N^{-0.58(1)}$.}
\end{figure}

The rapid changes observed in $\langle U(T) \rangle$ may suggest the possibility of a phase transition.
To find out if this is actually the case, we use the fourth-order Binder cumulant $K$ defined in \Cref{eq:K}.
Since we already know that the temperature at which the change occurs depends on the density of the graph, we need to find $K$ as a function of both $T$ and $D$.
The result obtained for $N=200$ is presented in \Cref{fig:K-D-T}.
It shows a distinct narrow minimum of $K$, which indicates the first-order phase transition \cite{Guide_to_Monte_Carlo_Simulations_2009,PhysRevB.34.1841,PhysRevE.59.218} and can be used to determine the critical temperature as a function of $D$.
However, for $N=200$ such a minimum of $K$ is observed only for densities greater than $D_{\text{min}} \approx 0.12$ (see the inset of \Cref{fig:K-D-T} using a finer mesh for calculations at low $D$ and $T$).
Simulations performed for different system sizes reveal that the minimum density required to observe the phase transition varies exponentially with $N$ as $D_{\text{min}} \propto N^{-0.58(1)}$, as demonstrated in \Cref{fig:Dmin}.
The data points presented in that figure were obtained from the arbitrarily chosen condition that the Binder cumulant $K$ becomes negative.

\begin{figure}
    \includegraphics[width=0.99\columnwidth]{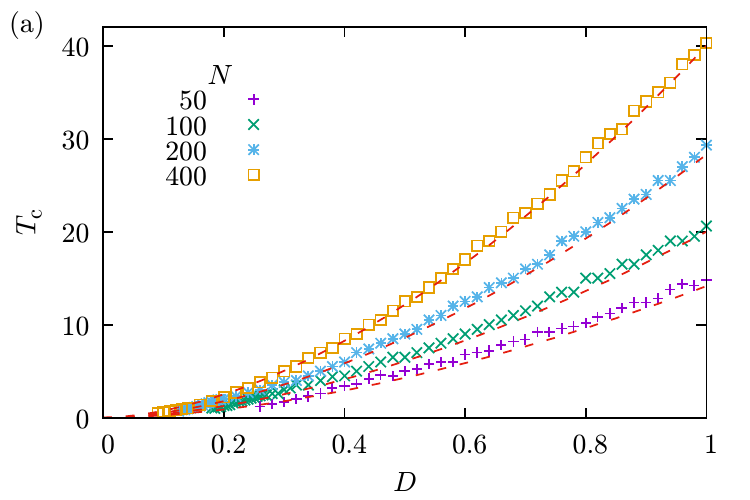}
    \includegraphics[width=0.99\columnwidth]{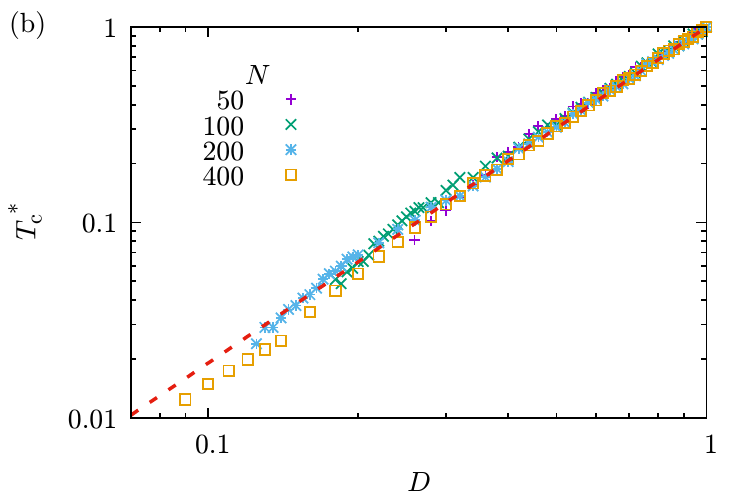}
    \caption{\label{fig:TC}(a) Critical temperature $T_\text{c}$ found from the minimum of the fourth-order Binder cumulant $K$ for $N=50, 100, 200$ and $400$. Dashed line used for $2 \sqrt{N} D^{1.719(6)}$.
    (b) Reduced critical temperature $T_\text{c}^\star = T_\text{c}/T_\text{c}(D=1)$ for the same system sizes, with dashed line showing the least-squares fitted $T_\text{c}^\star = D^{1.719(6)}$.}
\end{figure}

The values of $D_{\text{min}}$ are also the leftmost points of each of the data series (corresponding to various sizes $N=50, 100, 200$ and $400$) presented in \Cref{fig:TC}(a). 
It shows the critical temperatures $T_{\text{c}}$ that separate the balanced and imbalanced phases, determined as the minima of $K(T)$ for consecutive density values $D$.
The data in \Cref{fig:TC}(a) indicate that 
for all the analyzed system sizes $T_{\text{c}}$ increases with the graph density $D$.
Furthermore, for any given $D$, the value of $T_{\text{c}}$ increases with the size of the system.
The values of $T_{\text{c}}(D \to 1)\propto\sqrt N$ approach critical temperatures already known numerically for complete graphs (see Ref.~\onlinecite[Equation (2), Figure 2(b)]{2009.10136}).
Our results indicate that this $\sqrt{N}$-proportionality is preserved throughout the entire range of densities $D$, as shown by the dashed lines in \Cref{fig:TC}(a).
To establish a universal size-independent relation normalized to the value at density $D=1$, the reduced critical temperature defined as $T_\text{c}^\star = T_\text{c}/T_\text{c}(D=1)$ is calculated.
\Cref{fig:TC}(b) shows that this reduced critical temperature increases with the graph density as $T_\text{c}^\star = D^\gamma$, with $\gamma = 1.719(6)$ found using the least-squares linear fit for data in logarithmic scale.

\section{Discussion\label{sec:discussion}}

The observation that the critical temperature increases with the graph density $D$, but the phase transition occurs only if $D > D_{\text{min}}$, suggests that the internal structure of the system, i.e., the properties of the graph varying with $D$ may be related to the possibility of the phase transition.
Since the dynamics in the considered model is based on the relations in triads, it is important to note that at low densities relatively few triads exist in the graph.
This means that they tend to be separated with regard to the possible mutual influence on their states, which occurs only if the triads include a common relation.
The lack of interaction between at least some of the triads in the system can be analyzed in terms of the number of clusters composed of triads which may interact at least indirectly, with no interaction between triads belonging to different clusters.

\begin{figure}
    \includegraphics[width=0.99\columnwidth]{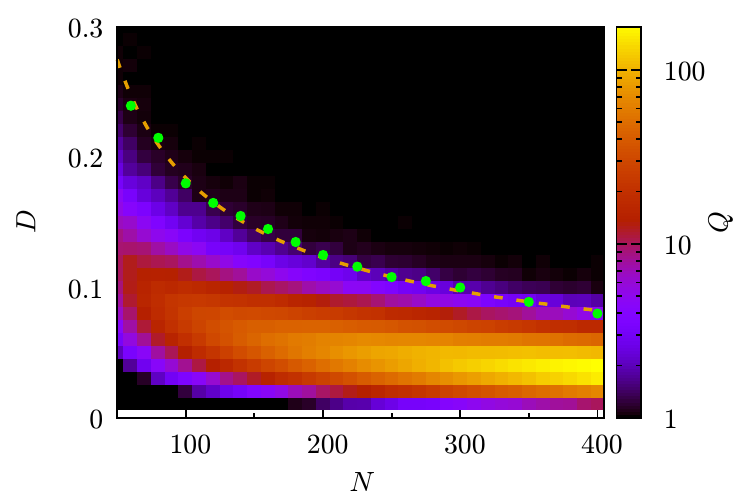}
    \caption{\label{fig:clusters}The number of clusters of interacting triads, $Q$, as a function o the system size $N$ and graph density $D$. Green dots represent the values of minimal density $D_{\text{min}}$ allowing the phase transition (see \Cref{fig:Dmin}).}
\end{figure}

To find out how the value of $D_{\min}$ may be related to the internal structure of the system, the number $Q$ of clusters of interacting triads was calculated using a version of the Hoshen--Kopelman algorithm\cite{Hoshen1976a,*Frijters_2015,*1803.09504} (see also Ref.~\onlinecite[pp.~59--60]{Guide_to_Monte_Carlo_Simulations_2009}) adapted to the situation when neighboring triads are defined as those that have a common link.
The results obtained as a function of the system size $N$ and density $D$ are presented in \Cref{fig:clusters}.
For a given $N$, the two limiting cases of low and high $D$ are self-evident. 
At the lowest densities, there is only a very limited number of isolated triads, each of them constituting its own cluster of one element, and hence the resulting low $Q$. 
The increase in $D$ adds more and more triads to the system, but initially with a low probability of containing a relation that also belongs to another triad.
As a result, the number of clusters quickly increases.
Then, when high densities are reached, the number of triads becomes very large, $\Delta = \Delta_{\textsc{cg}} D^3$, where $\Delta_{\textsc{cg}}=\binom{N}{3}$ is the number of triads in a complete graph with $N$ nodes; obviously, it quickly reduces the number of clusters to only one which can be treated as a percolation-like transition, however in this case not related to existence of clusters of nodes, but clusters of interacting triads.

The values of $D_{\text{min}}$ included in \Cref{fig:clusters} as green dots show that the phase transition occurs only if all triads belong to a limited number of large clusters.
For smaller sizes, it comes down to just one cluster with all triads, which enables the phase transition, while for the larger systems, the density must reach a value for which the number of clusters of interacting triads is reduced to just a few.
It seems to be the property of classical random graphs that the phase transition between balanced and imbalanced states is possible only if $Q$ is a small number.
However, when systems based on other types of graphs are considered, having just one cluster that contains all triads does not guarantee that the phase transition is achievable, as illustrated by the results of simulations performed on the regular triangular lattice~\cite{2106.03054}.

\begin{figure*}
    \includegraphics[width=0.99\textwidth]{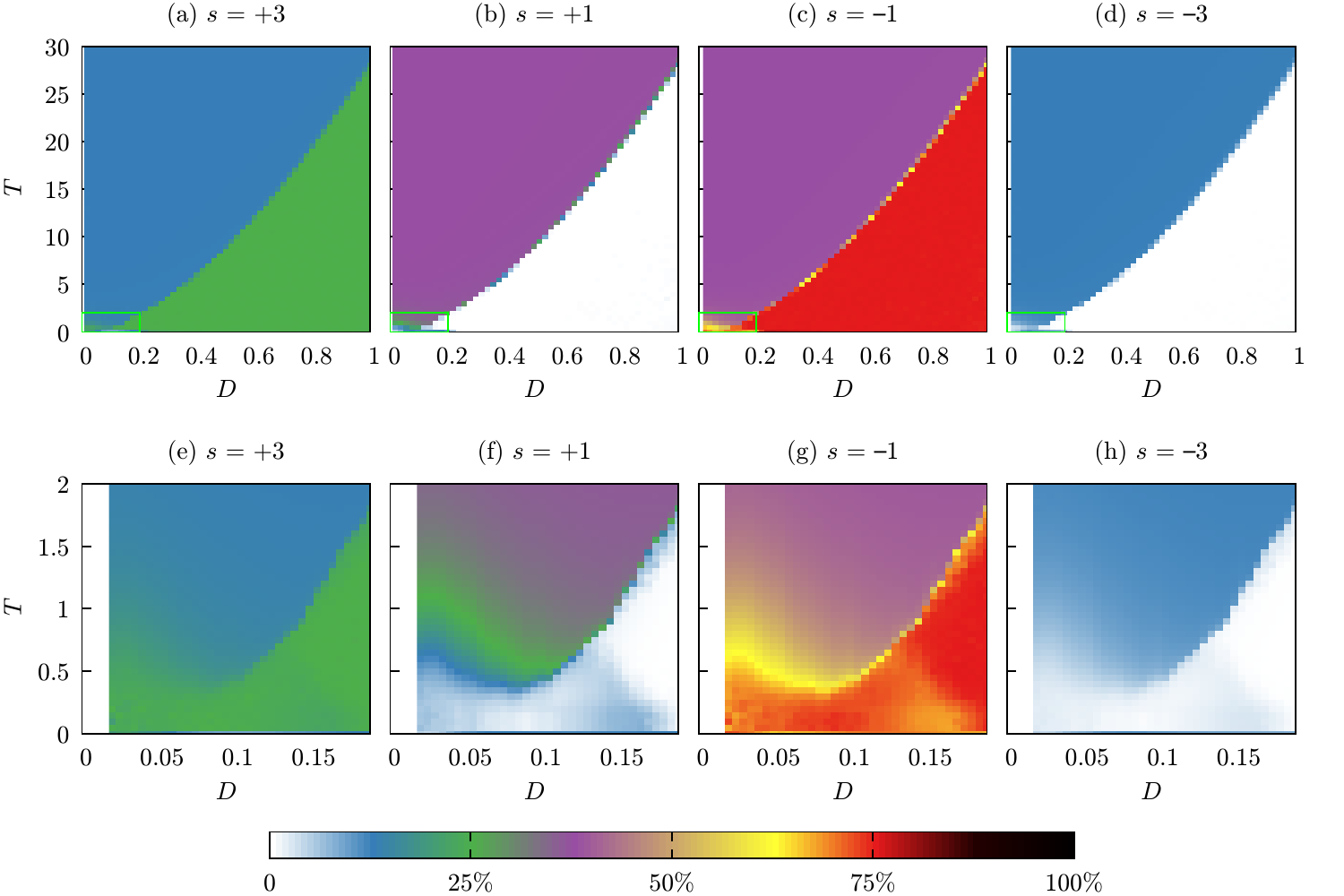}
    \caption{Dependence of the fraction of balanced [(a, e) $s=+3$ and (c, g) $s=-1$] and imbalanced [(b, f) $s=+1$ and (d, h) $s=-3$] triads on density $D$ and temperature $T$ calculated for $N=200$. (e--h) Detailed view for the range of low $D$ and low $T$ indicated with green rectangles in (a--d).}
    \label{fig:triads_vs_D_T}
\end{figure*}

The fact that for densities below $D_{\text{min}}$ the phase transition is not observed does not mean that the system remains in an imbalanced state when the temperature decreases.
As shown in \Cref{fig:crg-U_vs_T}(a) for $D=0.05$, $0.10$ and $0.15$ ($D_{\text{min}} \approx 0.18$ for $N=100$) or in \Cref{fig:crg-U_vs_T}(c) for $D=0.05$ ($D_{\text{min}} \approx 0.08$ for $N=400$), a partially balanced state with $\langle U (T \to 0 \rangle \approx -0.9$ is reached.
In this case, the system work function gradually changes with $T$ during a smooth crossover from the imbalanced state.

The difference between the two regimes, above and below $D_{\text{min}}$, is also visible in the proportion of the triad types in the system presented for $N=200$ in \Cref{fig:triads_vs_D_T} (at this size, the minimum density needed for the phase transition is $D_{\text{min}} \approx 0.12$).
In the first case, that is, for $D > D_{\text{min}}$, all triads are balanced when $T < T_{\text{c}}$; most of them (75\%) with $s=-1$, and the rest (25\%) with $s=+3$.
For a complete graph ($D=1$) this ratio (75\%$\div$25\%) can be calculated in the thermodynamic limit ($N\to\infty$) based on the assumption that the system is divided into two equinumerous cliques: internally friendly triangles [$s=+3$, see \Cref{fig:triads}(a)] connected by hostile triangles [$s=-1$, see \Cref{fig:triads}(c)] \cite{1911.13048}.
When the temperature increases, these proportions change rapidly at $T_{\text{c}}$ to 12.5\% for each of the $s=+3$ and $s=-3$ triads, and 37.5\% for each of the $s=+1$ and $s=-1$ triads, which is a simple consequence of all links $x_{ij}$ taking random ($\pm 1$) and independent values.
The second case of $D < D_{\text{min}}$, shown in detail in \Cref{fig:triads_vs_D_T}(e--h), is related to a smooth change in the proportion of triads and a lack of complete balance that manifests itself by a non-zero proportion of $s=+1$ and $s=-3$ triads even at $T \to 0$.
For $D$ just above $D_{\text{min}}$, the system reaches $\langle U \rangle = -1$ for temperatures in a certain range below $T_{\text{c}}$, but apparently this is not the case at the lowest temperatures.
An example of such a situation is discussed in \Cref{sec:results} for the case of $D = 0.15$ and $N=200$ [\Cref{fig:crg-U_vs_T}(b)].

\begin{figure}
\includegraphics[width=0.99\columnwidth]{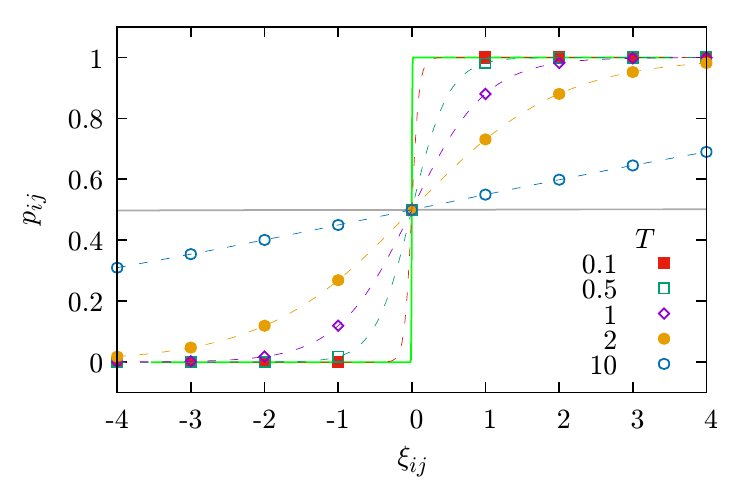}
\caption{\label{fig:pij_vs_xi}The probability $p_{ij}(\xi_{ij})$ [\Cref{eq:evol_p}] of setting a positive value for a link [\Cref{eq:evol_x}] for various $T$. 
Solid grey and green lines correspond to the cases of $T\to 0^+$ \eqref{eq:pij_limit_zero} and $T \to \infty$ \eqref{eq:pij_limit_infty}, respectively. Dashed lines are just guides for the eye. 
}
\end{figure}

\begin{figure*}
\includegraphics[width=\textwidth]{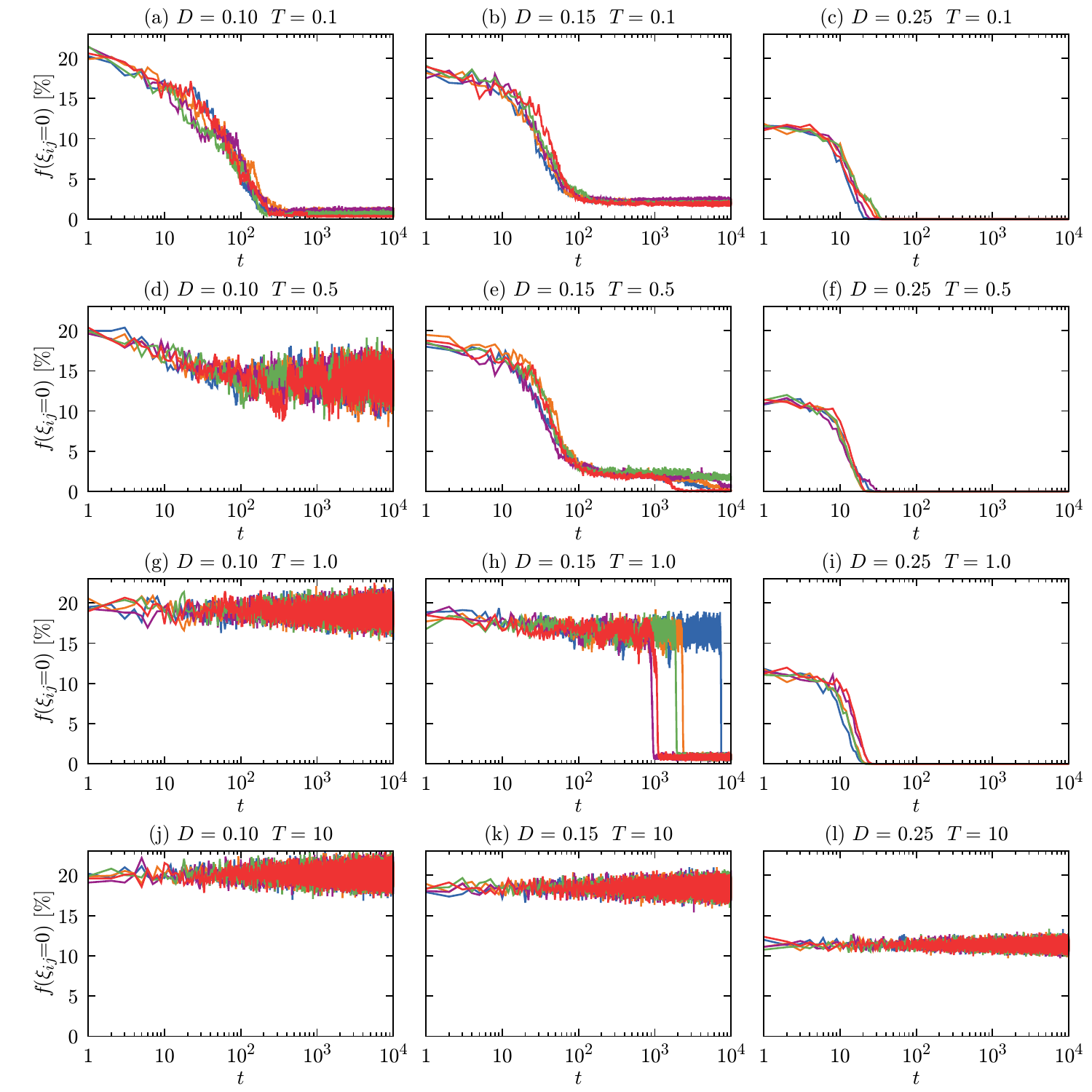}
\caption{\label{fig:crg-Ksi0_vs_time}(Color online). Time evolution of the percentage of links with $\xi_{ij}=0$ that belong to at least one triad, for systems described by classical random graphs of size $N=200$ with
various densities $D=0.10$, $0.15$, $0.25$, and temperatures $T=0.1$, $0.5$, $1$ and $10$. Results obtained in five independent simulations.}
\end{figure*}

However, even the time dependence of $U$ for those parameters, presented in \Cref{fig:crg-U_vs_time}(b, e, h), suggests that very long times are required to reach the final state, possibly much longer than reasonably possible during numerical simulations requiring collecting large statistics for a wide range of parameters.

Interestingly, this behavior is observed only for some intermediate values of the graph density and can be explained as follows.
In the low-temperature limit, the probability of establishing a positive value for a link---as predicted by \Cref{eq:evol_p}---is
\begin{equation}
\label{eq:pij_limit_zero}
\lim_{T\to 0^+} 
p_{ij}(t)=
        \begin{cases}
	1   & \text{ when }\xi_{ij}(t) > 0,\\
	1/2 & \text{ when }\xi_{ij}(t) = 0,\\
    0   & \text{ when }\xi_{ij}(t) < 0.
        \end{cases}
\end{equation}
The shape of the $p_{ij}(\xi_{ij})$ dependence is presented in \Cref{fig:pij_vs_xi} and for the low-temperature limit it tends to the Heaviside step function (see also Figure~2 in Reference~\onlinecite{2008.06362}). In other words, except for $\xi_{ij}=0$, the time evolution of the system promotes a paradise state as the final state of the system. 
The latter should result in $\langle U\rangle=-1$, which is observed for large enough ($D>0.25$) graph densities (see \Cref{fig:crg-U_vs_T}).
Yet the cases of $\xi_{ij}=0$ may be quite frequent and prevent reaching the paradise state.
As shown in \Cref{fig:crg-Ksi0_vs_time} it is especially true for lower densities, which coincides with the increasing number of clusters of interacting triads (see also \Cref{fig:clusters}).

\Cref{fig:triads_vs_D_T}(f) shows that this situation leads to a changed proportion of triads, with approximately 10\% of $s=+1$ triads (not observed in the balanced state) existing at $T \to 0$.
The numbers of other types of triads are also slightly changed compared to the balanced state reached by the system at higher $T$, as long as the temperature is still below $T_{\text{c}}$.
As a result, after finite simulation time $t_{\text{max}}$ the partial balance region spreads to higher densities at low $T$, compared to higher temperatures.

\section{Conclusions}

In this paper the possibility of reaching the Heider balance on Erd{\H o}s--R\'enyi graphs is studied by means of computer simulations based on the heat-bath algorithm.
The meaning of the steering parameter $T$ (thermal noise) in the heat-bath algorithm definition is not dissimilar to the social temperature (as defined in Refs.~\onlinecite{Microsociology,Macrosociology}) or information noise (as defined in opinion formation models \cite{1902.03454,*2002.05451,*2211.04183}): In the limit of $T\to\infty$ the relations are taken randomly
\begin{equation}
\label{eq:pij_limit_infty}
\forall \xi_{ij}\in \mathbb{Z} \lim_{T\to\infty} p_{ij}(t)=	1/2,
\end{equation}
while for $T\to 0^+$ the balanced triangles form clusters (also with only friendly relations---i.e. the paradise) which are stable (see Fig.~\ref{fig:pij_vs_xi} and Eq.~\eqref{eq:pij_limit_zero}).

The obtained results allowed for identification of the critical thermal noise $T_{\text{c}}$ separating globally balanced and imbalanced states and for observing the change of kind of phase transition from continuous (for low graph density) to abrupt (for higher density). Such change of order of the phase transition with network connectivity was observed earlier in opinion formation models \cite{Encinas_2018,*PhysRevE.106.014125}. For model were both opinion and relations are involved in construction of Hamiltonian the critical network density (for fixed social temperature) exists\cite{Pham_2020}. \citeauthor{Pham_2020} emphasise: `if the connectivity increases above the critical value, society must fragment'.

The structural balance in systems with an internal network of relations based on classical random graphs can be reached at least to some extent independently of the graph density $D$, provided that the level of thermal noise $T$ is sufficiently low.
It is remarkably different from the results obtained for the regular triangular lattice and its densified or diluted versions.
In the presence of thermal noise, even partial Heider balance is impossible not only in the ideal triangular lattices~\cite{2007.02128}, but also if they are randomly modified by adding or removing some links, provided that the resulting graph density is not too far from that characteristic to the regular grid~\cite{2106.03054}.

However, the density of a classical random graph has a decisive impact on how the transition between balanced and imbalanced states proceeds.
Our results show that above a certain minimal density, a phase transition of the first kind is observed, with the critical temperature dependent on the system size.
It is possible, nonetheless, to define a reduced critical temperature that is size-independent and scales with size as $T_{\text{c}}^\star\propto D^{1.719(6)}$.
Below the minimal density required for the phase transition, a gradual and smooth crossover between balanced and imbalanced states occurs.
It turns out that the minimal density mentioned above also varies with the size of the system, changing exponentially as $D_\text{min}\propto N^{-0.58(1)}$.
The presence of the phase transition is directly related to the number of clusters of triads that may interact (at least indirectly) during time evolution: we show that it is possible only when the system consists of a large enough number of triads organized into just one or a small number of such clusters.

\begin{acknowledgments}
The authors are grateful to Krzysztof Ku{\l}akowski for fruitful discussion and critical reading of the manuscript.
\end{acknowledgments} 

\bibliography{bib/km,bib/heider,bib/basics,bib/networks,bib/opiniondynamics,bib/percolation,bib/this}

\end{document}